\documentclass[twocolumn,showpacs,preprintnumbers,pre]{revtex4}
\usepackage{epsfig,bm}
\usepackage{graphicx}
\usepackage{dcolumn}
\usepackage{bm}
\begin{document}

\title{SOME NON-PERTURBATIVE AND NON-LINEAR EFFECTS IN LASER-ATOM
INTERACTION}

\author{Qi-Ren Zhang
\\CCAST(World Lab),P.O.Box 8730,Beijing,100080, China\\and
\\Department of Technical Physics, Peking University ,
Beijing,100871,China}

\vskip0.3cm

\begin{abstract}
We show that if the laser is intense enough, it may always ionize an
atom or induce transitions between discrete energy levels of the
atom, no matter what is its frequency. It means in the quantum
transition of an atom interacting with an intense laser of circular
frequency $\omega$, the energy difference between the initial and
the final states of the atom is not necessarily being an integer
multiple of the quantum energy $\hbar\omega$. The absorption spectra
become continuous. The Bohr condition is violated. The energy of
photoelectrons becomes light intensity dependent in the intense
laser photoelectric effect. The transition probabilities and cross
sections of photo-excitations and photo-ionizations are laser
intensity dependent, showing that these processes cannot be reduced
to the results of interactions between the atom and separate
individual photons, they are rather the processes of the atom
interacting with the laser as a whole. The interaction of photons on
atoms are not simply additive. The effects are non-perturbative and
non-linear. Some numerical results for processes between hydrogen
atom and intense circularly polarized laser, illustrating the
non-perturbative and non-linear character of the atom-laser
interaction, are given.


\end{abstract}


\maketitle

\section{Introduction}
From the matrix element of electromagnetic interaction Hamiltonian
we see that the strength of this interaction is characterized by
$\mathcal {N}\alpha$, in which $\alpha$ is the fine structure
constant, and $\mathcal {N}$ is the number of photons in the
electromagnetic wave mode participating the interaction. In the
processes with usual light, $\mathcal {N}\sim 1$, the strength of
the electromagnetic interaction is of the order of
$\alpha=1/137.0<<1$, it is weak, and therefore may be handled by
lowest order perturbation. However, in the processes with lasers,
$\mathcal {N}$ becomes large, the electromagnetic interaction is no
longer weak. In this case, either a high order perturbation has to
be used, or perturbation becomes totally unapplicable. Anyway, the
process induced by intense laser with huge number $\mathcal {N}$ can
only be handled by non-perturbation method. To explore the
non-perturbative effects in electromagnetic processes is
interesting.

Soon after the discovery of laser, Voronov and Delone\cite{v}
observed the multi-photon ionization (MPI), in which an atom absorbs
more than one photons from the laser and is ionized. Several years
later, Agostini {\it et al}\cite{a} observed the above threshold
ionization (ATI). It is the photo-ionization with the photoelectron
energy being larger than the photon energy. When the laser intensity
is not too high, they may still be understood by
perturbation\cite{g}, though the perturbation order has to be high.
In 1988, Bucksbaum {\it et al}\cite{B} observed electrons ejected
from atoms and scattered to large angles by an intense standing
electromagnetic wave, which is something like the Kapitza-Dirac
effect\cite{k}. The angular distribution of electron in this
experiment is very characteristic. Guo and Drake\cite{g0} explained
Bucksbaum's experiment non-perturbatively using a theory developed
by Guo {\it et al}\cite{ga} on the basis of KFR
theory\cite{ke}-\cite{r} for atom-laser interaction.
Batelaan\cite{ba,fb} observed the Kapitza-Dirac effect in 2000. It
is of course a non-perturbative effect in electron-laser
interaction. The position of peaks in electron diffraction may be
well understood by wave property of electrons and geometrical
consideration. To understand the height of these diffraction peaks
one needs the quantum dynamics for electron-laser system. We
showed\cite{xz} the excellent agreement between the quantum
dynamical calculation and the experiment in this respect.

The main problem in theoretical consideration of the laser-atom
interaction is to solve the electron motion under the combined
interaction of the Coulomb field and the laser field. To find an
analytical solution, even though approximate, seems hopeless.
Another choice is the numerical computation. Some people tried to
solve the time-dependent Schr\"{o}dinger equation
numerically\cite{l,y}. To achieve an acceptable numerical solution,
the amount of computation is tremendous. The problem becomes how to
control the amount of numerical computation within a reasonable
limit, so that the computation may be realized and the accumulative
error remains acceptable. It is to find a balance between the
analytical derivation and the numerical computation.

There is a way to solve the time dependence of the Schr\"{o}dinger
equation analytically for the atom in a circularly polarized
laser\cite{z}. In this way the problem of the atom-laser interaction
reduces to an eigenvalue problem of a time independent effective
Hamiltonian for the atom. The energy representation of the effective
Hamiltonian is a matrix of order infinite. In numerical solution,
one has to truncate it, to make it be of finite order. Then solve
the eigenvalue problem of the finite order matrix, and search the
limit of its solution when the order of the matrix approaches
infinite.  We found the rapid convergence of the solution when the
order of matrix increases\cite{zz}. The problem of hydrogen in a
circularly polarized laser may therefore be solved by a reasonable
amount of numerical computation.

We quantize the electromagnetic field around a classical field
representing the laser in section \ref{s2}, to separate the
perturbation and non-perturbation parts of the problem. In section
\ref{s3}, we solve the time dependence of the Schr\"odinger equation
analytically and separate the relative motion between the electron
and proton from the motion of center of mass for a non-relativistic
hydrogen atom interacting with a circularly polarized laser. In the
dipole limit, the modification of the nucleus motion on the relative
motion is again the substitution of reduced mass for the electron
mass. We solve the transition between discrete levels and the
photo-ionization of hydrogen atom irradiated by circularly polarized
laser numerically in sections \ref{s4} and \ref{s5} respectively.
Non-perturbative and non-linear effects are shown in figures.
Section \ref{s7} is the conclusion.

\section{Quantization of the electromagnetic field around a classical field
and the separation of the non-perturbative and perturbative parts of
the problem\label{s2}} In the Coulomb gauge, the quantization of the
electromagnetic field is to let the vector potential be an operator
$\hat{\textit{\textbf{A}}}$ and define commutators between its
components. Introducing a complete set of vector functions
$[{\textit{\textbf{A}}}_\iota({\textit{\textbf{r}}})]$, satisfying
the Helmholtz equations
\begin{eqnarray}
\nabla^2{\textit{\textbf{A}}}_\iota({\textit{\textbf{r}}})
+\frac{\omega_\iota^2}{c^2}{\textit{\textbf{A}}}_\iota({\textit{\textbf{r}}})=0
\end{eqnarray}
and the orthonomal conditions
\begin{eqnarray}
\int{\textit{\textbf{A}}}_\iota^*({\textit{\textbf{r}}})\cdot
{\textit{\textbf{A}}}_\iota^\prime({\textit{\textbf{r}}}){\rm
d}{\textit{\textbf{r}}} = \delta_{\iota\iota^\prime}\ ,
\end{eqnarray}
one may expand the self-adjoint operator
\begin{eqnarray}
\hat{\textit{\textbf{A}}}({\textit{\textbf{r}}},t)
=\sum_\iota\sqrt{\frac{\hbar}{2\epsilon_0\omega_\iota}}
\left[\hat{b}_\iota{\textit{\textbf{A}}}_\iota({\textit{\textbf{r}}})
+\hat{b}_\iota^\dag{\textit{\textbf{A}}}_\iota^*({\textit{\textbf{r}}})
\right]\ ,
\end{eqnarray}
$\epsilon_0$ is the dielectric constant for the vacuum. The
quantization condition is the commutators
\begin{eqnarray}
\hat{b}_\iota\hat{b}_{\iota^\prime}-\hat{b}_{\iota^\prime}\hat{b}_\iota
&=&\hat{b}_\iota^\dag\hat{b}_{\iota^\prime}^\dag-\hat{b}_{\iota^\prime}^\dag
\hat{b}_\iota^\dag=0,\;\;\;\;\mbox{and}\nonumber\\
\hat{b}_\iota\hat{b}_{\iota^\prime}^\dag-\hat{b}_{\iota^\prime}^\dag
\hat{b}_\iota&=&\delta_{\iota\iota^\prime}\ .
\end{eqnarray}
The vacuum state $|0\rangle$ is defined by
\begin{eqnarray}
\hat{b}_\iota|0\rangle=0\;\;\;\;\;\;\mbox{for all $\iota$}\ .
\end{eqnarray}
This is the quantization around the vacuum. Classically, the vacuum
is described by a vector potential
${\textit{\textbf{A}}}_0({\textit{\textbf{r}}},t)=0$, which is a
trivial solution of the D'Alembert equation. It suggests, that one
may also quantize the theory around another classical solution
${\textit{\textbf{A}}}_c({\textit{\textbf{r}}},t)$ of the D'Alembert
equation, for example the circularly polarized plane wave
\begin{eqnarray}
{\textit{\textbf{A}}}_c=A[{\textit{\textbf{x}}}_0\cos(kz-{\omega}t)
+{\textit{\textbf{y}}}_0\sin(kz-{\omega}t)]\  .\label{11}
\end{eqnarray}
Defining $\hat{\textit{\textbf{A}}}^\prime({\textit{\textbf{r}}},t)
=\hat{\textit{\textbf{A}}}({\textit{\textbf{r}}},t)
-{\textit{\textbf{A}}}_c({\textit{\textbf{r}}},t)$, expanding
\begin{eqnarray}
{\textit{\textbf{A}}}_c({\textit{\textbf{r}}},t)
=\sum_\iota\sqrt{\frac{\hbar}{2\epsilon_0\omega_\iota}}
\left[c_\iota{\textit{\textbf{A}}}_\iota({\textit{\textbf{r}}})
+c_\iota^*{\textit{\textbf{A}}}_\iota^*({\textit{\textbf{r}}})
\right]\ ,
\end{eqnarray}
we have
\begin{eqnarray}
\hat{\textit{\textbf{A}}}^\prime({\textit{\textbf{r}}},t)
=\sum_\iota\sqrt{\frac{\hbar}{2\epsilon_0\omega_\iota}}
\left[\hat{b}_\iota^\prime{\textit{\textbf{A}}}_\iota({\textit{\textbf{r}}})
+\hat{b}_\iota^{\prime\dag}{\textit{\textbf{A}}}_\iota^*({\textit{\textbf{r}}})
\right]\ ,
\end{eqnarray}
with
\begin{eqnarray}
\hat{b}^\prime_\iota=\hat{b}_\iota-c_\iota\ .
\end{eqnarray}
Since $c_\iota$ are c-numbers, $\hat{b}_\iota^\prime$ and
$\hat{b}_\iota^{\prime\dag}$ have the same commutators as those for
$\hat{b}_\iota$ and $\hat{b}_\iota^\dag$. They are
\begin{eqnarray}
\hat{b}_\iota^\prime\hat{b}_{\iota^\prime}^\prime-\hat{b}_{\iota^\prime}^\prime\hat{b}_\iota^\prime
&=&\hat{b}_\iota^{\prime\dag}\hat{b}_{\iota^\prime}^{\prime\dag}-\hat{b}_{\iota^\prime}^{\prime\dag}
\hat{b}_\iota^{\prime\dag}=0,\;\;\;\;\mbox{and}\nonumber\\
\hat{b}_\iota^\prime\hat{b}_{\iota^\prime}^{\prime\dag}-\hat{b}_{\iota^\prime}^{\prime\dag}
\hat{b}_\iota^\prime&=&\delta_{\iota\iota^\prime}\ .
\end{eqnarray}
The quantization condition for the electromagnetic field
$\hat{\textit{\textbf{A}}}^\prime$ around a classical field
${\textit{\textbf{A}}}_c$ is therefore of the same form as that for
the field $\hat{\textit{\textbf{A}}}$ around the classical vacuum
${\textit{\textbf{A}}}_0=0$. However, the 'vacuum' state is now
changed to $|c\rangle$, satisfying
$\hat{b}_\iota^\prime|c\rangle=0$. This is
\begin{eqnarray}
\hat{b}_\iota|c\rangle=c_\iota|c\rangle\ ,
\end{eqnarray}
showing that $|c\rangle$ is a coherent state with non-zero
amplitude(s) $c_\iota$. In the classical limit it is
${\textit{\textbf{A}}}_c$ itself.

The interaction operator between the electromagnetic field and a
non-relativistic spinless electron is
\begin{eqnarray}
\frac{e}{m}\hat{\textit{\textbf{A}}}\cdot\hat{\textit{\textbf{p}}}
&+&\frac{e^2\hat{\textit{\textbf{A}}}^2}{2m}=\frac{e}{m}{\textit{\textbf{A}}}_c
\cdot\hat{\textit{\textbf{p}}}+\frac{e^2A_c^2}{2m}\nonumber\\
&+&\frac{e}{m}\hat{\textit{\textbf{A}}}^\prime\cdot\hat{\textit{\textbf{p}}}
+\frac{e^2\hat{\textit{\textbf{A}}}^{\prime2}}{2m}+\frac{e^2
\hat{\textit{\textbf{A}}}^\prime\cdot{\textit{\textbf{A}}}_c}{m}\ ,
\end{eqnarray}
in which $-e$, $m$ and $\textit{\textbf{p}}$ are charge, mass, and
momentum respectively of the electron. If ${\textit{\textbf{A}}}_c$
represents an intense laser, the first two terms on the right hand
side would be large, its effect has to be treated
non-perturbatively. It is the main part of the problem, and is
exactly the interaction in the semiclassical theory for the
atom-laser system. For small fluctuations of the electromagnetic
field around the laser, the number $\mathcal {N}\, ^\prime$ of
photons other than those in the laser is few, the matrix element of
the remaining terms on the right of this equation are small, and may
be considered by perturbation. The non-perturbative part and the
perturbative part of the problem are therefore separated.

\section{Solution of the time dependence and the separation of the
relative motion \label{s3} } Consider the atom-laser processes by
the semiclassical theory. According to the analysis in above
section, it is the main part of the physics for the atom-laser
system. The hydrogen atom consists of a proton and an electron. Its
Hamiltonian in the circularly polarized electromagnetic wave
(\ref{11}) is
\begin{eqnarray}
\hat{H}=\hat{H}_0+\hat{H}^\prime \ ,\label{12}
\end{eqnarray}
with
\begin{eqnarray}
\hat{H}_0=\frac{\hat{p}_1^2}{2m_1}+\frac{\hat{p}_2^2}{2m_2}+V(r)\
,\label{130}
\end{eqnarray}
and
\begin{eqnarray}
&&\hat{H}^\prime  =\frac{eA}{m_1} [\hat{p}_{1x}\cos(kz_1-\omega
t)+\hat{p}_{1y}\sin(kz_1-\omega t)]+\frac{e^2A^2}{2m_1}\nonumber
\\&-&\frac{eA}{m_2}[\hat{p}_{2x}\cos(kz_2-\omega
t)+\hat{p}_{2y}\sin(kz_2-\omega t)]+\frac{e^2A^2}{2m_2} .\label{140}
\end{eqnarray}
Subscripts 1 and 2 denote the electron and the proton respectively,
$r$ is the distance and $V(r)=-\frac{\hbar c\alpha}{r}$ is the
Coulomb potential between them. (\ref{140}) makes the Hamiltonian
(\ref{12}) and the corresponding Schr\"odinger equation
\begin{eqnarray}
{\rm i}\hbar\frac{\partial\Psi}{\partial t}=\hat{H}\Psi \label{15}
\end{eqnarray}
time dependent. However the transformation
\begin{eqnarray}
\Psi({\textit{\textbf r}}_1,{\textit{\textbf r}}_2,t)={\rm e}^{{\rm
i}\omega(\hat{L}_{1z}+\hat{L}_{2z})t /\hbar}\Phi({\textit{\textbf
r}}_1{\textit{\textbf r}}_2,t) \label{160}
\end{eqnarray}
changes the equation into a time independent pseudo-Schr\"odinger
equation
\begin{eqnarray}
{\rm i}\hbar\frac{\partial\Phi}{\partial t}=\hat{H}_{\rm ps}\Phi\ ,
\label{17}
\end{eqnarray}
with the pseudo-Hamiltonian
\begin{eqnarray}
\hat{H}_{\rm ps}=\hat{H}_0^\prime +\hat{H}^{\prime\prime}\
,\label{18}
\end{eqnarray}
in which
\begin{eqnarray}
\hat{H}_0^\prime=\hat{H}_0+\omega\sum_{j=1}^2\hat{L}_{jz}\label{190}
\end{eqnarray}
and
\begin{eqnarray}
\hat{H}^{\prime\prime}&=&\frac{eA}{m_1}
[\hat{p}_{1x}\cos(kz_1)+\hat{p}_{1y}\sin(kz_1)]+\frac{e^2A^2}{2m_1}
\nonumber
\\&-&\frac{eA}{m_2}[\hat{p}_{2x}\cos(kz_2)+\hat{p}_{2y}\sin(kz_2)]+\frac{e^2A^2}{2m_2}
\label{200}
\end{eqnarray}
are time independent. The time dependence of the Hamiltonian
(\ref{12}) and the corresponding Schr\"odinger equation (\ref{15})
is solved analytical by the operator ${\rm e}^{{\rm
i}\omega(\hat{L}_{1z}+\hat{L}_{2z})t /\hbar}$ in transformation
(\ref{160}). A further transformation
\begin{eqnarray}
\Phi=\exp\left[-{\rm
i}k\frac{(m_1z_1+m_2z_2)(\hat{L}_{1z}+\hat{L}_{2z})}{(m_1+m_2)\hbar}\right]\Phi_{\rm
e}
\end{eqnarray}
changes (\ref{17}) into
\begin{eqnarray}
{\rm i}\hbar\frac{\partial\Phi_{\rm e}}{\partial t}=\hat{H}_{\rm
e}\Phi_{\rm e}\ , \label{170}
\end{eqnarray}
with the effective Hamiltonian
\begin{eqnarray}
&\hat{H}_{\rm
e}=\sum_{j=1}^2\left[\frac{\hat{p}_{jx}^2+\hat{p}_{jy}^2+(\hat{p}_{jz}
-m_jk\frac{\hat{L}_{1z}+\hat{L}_{2z}}{m_1+m_2})^2}{2m_j}
+\omega\hat{L}_{jz}\right]\nonumber\\&+V(r)+\frac{eA}{m_1}
\left[\hat{p}_{1x}\cos\left(km_2\frac{z_1-z_2}{m_1+m_2}\right)\right.\nonumber\\
&\left.
+\hat{p}_{1y}\sin\left(km_2\frac{z_1-z_2}{m_1+m_2}\right)\right]-\frac{eA}{m_2}
\left[\hat{p}_{2x}\cos\left(km_1\frac{z_2-z_1}{m_1+m_2}\right)\right.
\nonumber
\\&\left.
+\hat{p}_{2y}\sin\left(km_1\frac{z_2-z_1}{m_1+m_2}\right)\right]+\sum_{j=1}^2\frac{e^2A^2}{2m_j}
\ .\label{180}
\end{eqnarray}
Introducing the center of mass coordinates ${\textit{\textbf
R}}\equiv(m_1{\textit{\textbf r}}_1+m_2{\textit{\textbf r}}_2)/M$
and the relative coordinates ${\textit{\textbf
r}}\equiv{\textit{\textbf r}}_1-{\textit{\textbf r}}_2$,
 we have the total momentum
$\hat{\textit{\textbf P}}=\hat{\textit{\textbf
p}}_1+\hat{\textit{\textbf p}}_2$, the relative momentum
$\hat{\textit{\textbf p}}=m(\hat{\textit{\textbf
p}}_1/m_1-\hat{\textit{\textbf p}}_2/m_2)$, the angular momentum
$\hat{\textit{\textbf L}}_c={\textit{\textbf
R}}\times\hat{\textit{\textbf P}}$ of the center of mass, and the
angular momentum $\hat{\textit{\textbf L}}={\textit{\textbf
r}}\times\hat{\textit{\textbf p}}$ around the center of mass.
$M=m_1+m_2$ is the total mass, and $m=m_1m_2/(m_1+m_2)$ is the
reduced mass. In these coordinates, the effective Hamiltonian
(\ref{180}) has the form
\begin{eqnarray}
\hat{H}_{\rm e}&=&\frac{\hat{\textit{\textbf
P}}_x^2+\hat{\textit{\textbf P}}_y^2+[\hat{\textit{\textbf P}}_z
-k(\hat{L}_{cz}+\hat{L}_z)]^2}{2M} +\omega\hat{L}_{cz}\nonumber\\
&+&\frac{p^2}{2m}+V(r)+\omega\hat{L}_z+\frac{eA}{m}\hat{p}_x
+\frac{e^2A^2}{2m}\ ,\label{201}
\end{eqnarray}
at the dipole limit  $ka_0\rightarrow0 $, $a_0$ is the Bohr radius.
The sum of the last five terms relates to the relative motion only.
The first two terms mainly relate to the motion of the center of
mass. Only $\hat{L}_z$ in the first term relates to the relative
motion. But the big mass $M$ on the denominator makes its
contribution be much less than that of the last five terms.
Therefore one needs only to consider the sum of the last five terms
in (\ref{201}) for the problem of relative motion between the
electron and the proton in hydrogen atom, irradiated by a circularly
polarized laser. The first two terms in (\ref{201}) govern the
motion of the hydrogen atom as a whole. They have to be considered
if one is interested in details of the ionized electrons, for
example, in their angular distributions.

Since $m_2>>m_1$, we may simplify the problem by taking the limit
$m_2\rightarrow\infty$. It becomes a single body problem of electron
under the combined interaction of the Coulomb potential around a
fixed point and the laser field. Equations (\ref{130}), (\ref{140}),
(\ref{160}), (\ref{190}), and (\ref{200}) become
\begin{eqnarray}
\hat{H}_0=\frac{\hat{p}^2}{2m}+V(r)\ ,\label{13} \end{eqnarray}
\begin{eqnarray}
\hat{H}^\prime &=&\frac{eA}{m} [\hat{p}_x\cos(kz-\omega
t)+\hat{p}_y\sin(kz-\omega t)]+\frac{e^2A^2}{2m}\ ,\nonumber\\
\label{14}
\end{eqnarray}
\begin{eqnarray}
\Psi({\textit{\textbf r}},t)={\rm e}^{{\rm i}\omega\hat{L}_zt
/\hbar}\Phi({\textit{\textbf r}},t)\ ,\label{16}
\end{eqnarray}
\begin{eqnarray} \hat{H}_0^\prime=\hat{H}_0+\omega\hat{L}_z
\ ,\label{19}\end{eqnarray} and \begin{eqnarray}
\hat{H}^{\prime\prime}=\frac{eA}{m}
[\hat{p}_x\cos(kz)+\hat{p}_y\sin(kz)]+\frac{e^2A^2}{2m} \label{20}
\end{eqnarray}
respectively. The subscript 1 is omitted. At the dipole limit
$ka_0<<1$, we have (\ref{17}) with
\begin{eqnarray}
\hat{H}_{\rm
ps}=\frac{p^2}{2m}+V(r)+\omega\hat{L}_z+\frac{eA}{m}\hat{p}_x
+\frac{e^2A^2}{2m}\ .\label{202}
\end{eqnarray}
Comparing this equation with the last five terms in (\ref{201}) we
see, at the dipole limit, the influence of the nucleus motion on the
relative motion between the electron and nucleus is again the
substitution of the reduced mass for the electron mass. The
modification is tiny.

\section{Transitions between discrete levels of the hydrogen atom
irradiated by a circularly polarized laser\label{s4} } In the
following numerical computations the dipole limit condition is
always well satisfied. Our work is to solve the pseudo-Schr\"odinger
equation (\ref{17}) with time independent pseudo-Hamiltonian
(\ref{202}). Denote the $i$th eigenfunction of $\hat{H}_{\rm ps}$ by
$\phi_i({\textit{\textbf r}})$. We have
\begin{eqnarray}
\hat{H}_{\rm ps}\phi_i({\textit{\textbf
r}})=E_i\phi_i({\textit{\textbf r}})\ ,\label{21}
\end{eqnarray}
$E_i$ is the pseudo-energy of the electron in the
pseudo-stationary state $\phi_i({\textit{\textbf r}})$. They may
be quite different from the energy $E_n=-\alpha^2mc^2/2n^2$ and
the stationary state
\begin{eqnarray}
&&\psi_{nl\mu}({\textit{\textbf r}})=\frac{{\rm
i}^l}{(2l+1)!}\left[\left(\frac{1}{na_0}\right)^{2l+3}\frac{(n+l)!}{2n(n-l-1)!}\right]^{1/2}
\nonumber\\&&\times{\rm e}^{-\frac{r}{na_0}}r^l{\rm
F}(l+1-n,2l+2,\frac{2r}{na_0}){\rm
Y}_{l\mu}(\theta\varphi)\label{22}
\end{eqnarray}
of the electron in an isolated hydrogen atom respectively.  F is the
confluent hypergeometric function, and Y is the spherical harmonic
function. $r,\theta$ and $\varphi$ are spherical coordinates of the
electron. Now, let us expand $\phi_i$ in terms of the wave functions
[$\psi_{nl\mu}$]:
\begin{eqnarray}
\phi_i({\textit{\textbf
r}})=\sum_{nl\mu}C_{nl\mu}(i)\psi_{nl\mu}({\textit{\textbf r}})\
.\label{23}
\end{eqnarray}
This is an approximation, since the set [$\psi_{nl\mu}$] of bound
states only is not complete. We expect that it is good enough for
the state $\phi_i$ near a low lying  bound state. We further assume
that in the expansion (\ref{23}) only terms with $n\le n_0$ are
important, therefore one may truncate the summation on the right at
$n=n_0$. We saw fast convergence of the outcome with increasing
$n_0$ in numerical calculations\cite{zz}. We also see this kind of
convergence in the following numerical calculation. The truncation
makes the eigen-equation (\ref{21}) become algebraic, and therefore
may be solved by the standard method\cite{w}.

The factor ${\rm i}^l$ on the right of (\ref{22}) makes the matrix
elements of $\hat{H}_{\rm ps}$ be real in this representation.
Therefore the solutions $[C_{nl\mu}(i)]$ are real. We have the
reciprocal expansion
\begin{eqnarray}
\psi_{nl\mu}({\textit{\textbf
r}})=\sum_iC_{nl\mu}(i)\phi_i({\textit{\textbf r}})\ .\label{24}
\end{eqnarray}
Suppose the hydrogen atom stays in the state $\psi_{nl\mu}$ when
$t\leq 0$. The laser arrives at $t\!=\!0$. According to
(\ref{16}), (\ref{17}) and (\ref{21}), at $t>0$, the pseudo-state
will be
\begin{eqnarray}
&&\Phi({\textit{\textbf
r}},t)=\sum_iC_{nl\mu}(i)\phi_i({\textit{\textbf r}}){\rm e}^{-{\rm
i}E_it/\hbar}\nonumber\\&&=\sum_{n'l'\mu'}\sum_iC_{nl\mu}(i)C_{n'l'\mu'}(i){\rm
e}^{-{\rm i}E_it/\hbar}\psi_{n'l'\mu'}({\textit{\textbf r}})\ ,
\label{25}
\end{eqnarray}
and the state becomes
\begin{eqnarray}
\Psi({\textit{\textbf
r}},t)&=&\sum_{n'l'\mu'}\sum_iC_{nl\mu}(i)C_{n'l'\mu'}(i)\nonumber\\&\times&{\rm
e}^{-{\rm
i}(E_i-\mu'\hbar\omega)t/\hbar}\psi_{n'l'\mu'}({\textit{\textbf
r}}). \label{251}
\end{eqnarray}
The transition probability of the hydrogen atom from the state
$\psi_{nl\mu}$ to the state $\psi_{n'l'\mu'}$ is
\begin{eqnarray}
w_{n'l'\mu';nl\mu}(t)=\left| \sum_iC_{nl\mu}(i)C_{n'l'\mu'}(i){\rm
e}^{-{\rm i}(E_i-\mu'\hbar\omega)t/\hbar}\right|^2 .
\nonumber\\\label{26}
\end{eqnarray}
It is a multi-periodic function of $t$. The periods are of the
microscopic order of magnitude. On the other hand, the observation
is done in a macroscopic duration. Therefore the observed
transition probability is a time average of (\ref{26}) over its
periods. The averages of the cross terms with different $i$ in the
summation are zeros. It makes the observed transition probability
be
\begin{eqnarray}
W_{n'l'\mu':nl\mu}=\sum_iC_{nl\mu}^2(i)C_{n'l'\mu'}^2(i)\
.\label{27}
\end{eqnarray}
From the normalizations
\begin{eqnarray}
\sum_{nl\mu}C_{nl\mu}^2(i)=1\;\;\;\;\;\mbox{and}\;\;\;\;\;
\sum_iC_{nl\mu}^2(i)=1\label{28}
\end{eqnarray}
one sees the normalization
\begin{eqnarray}
\sum_{n'l'\mu'}W_{n'l'\mu';nl\mu}=1\ .
\end{eqnarray}
It shows that the expression (\ref{27}) for the transition
probability is reasonable.

When the amplitude $A$ is small (weak light), one may solve equation
(\ref{21}) by perturbation. The unperturbed pseudo-Hamiltonian is
$\hat {H}_0^\prime$, the unperturbed pseudo-states are
$[\psi_{nl\mu}]$, with unperturbed pseudo-energies
$[E_n+\mu\hbar\omega]$, and the perturbation is
\begin{eqnarray}
\hat{H}^{\prime\prime}=\frac{eA}{m}\hat{p}_x\ .\label{280}
\end{eqnarray}
The selection rules of its non-zero matrix elements include
\begin{eqnarray}
\Delta\mu\equiv\mu'-\mu=\pm 1\ .\label{281}
\end{eqnarray}
If the Bohr condition
\begin{eqnarray}
E_{n'}-E_n=\pm\hbar\omega \label{29}
\end{eqnarray}
is fulfilled, pseudo-states $\psi_{nl\mu}$ and $\psi_{n'l'\mu'}$
with $\Delta\mu=\pm 1$ are degenerate. The correct zeroth-order
approximation of eigen-states of $\hat{H}_{\rm ps}$ has to be formed
by their superpositions. The problem is equivalent to an eigenvalue
problem of a two level system. In the limit of $A\rightarrow 0$, a
resonance factor
\begin{eqnarray}
\lim_{t\rightarrow
\infty}\frac{\sin^2[(E_{n'}-E_n\mp\hbar\omega)\frac{t}{2\hbar}]}
{(E_{n'}-E_n\mp\hbar\omega)^2\frac{t}{2\hbar}}=\pi\delta(E_{n'}-E_n\mp\hbar\omega)\nonumber\\
\label{4}
\end{eqnarray}
appears. On the contrary, if the condition (\ref{29}) is not
fulfilled, the transition probability is zero in the zeroth-order
perturbation. We see, the transition probability calculated by
(\ref{27}) is  in agreement with that obtained by the traditional
method. This result may be regarded as a check of the method
proposed here. Now let us use it to consider the transitions in
lasers.

The $\hat{H}_0^\prime$ representation of $\hat{H}_{\rm ps}$, after
being truncated at $n_0=18$, is a $2109\times 2109$ matrix. It is
solved numerically by the standard method\cite{w} for various values
of $A$ and $k$.  Substituting the solved eigen-vectors into
(\ref{27}), we obtain transition probabilities for these cases. The
results are shown in the figures.

\begin{figure}[h]
\centerline{\epsfig{file=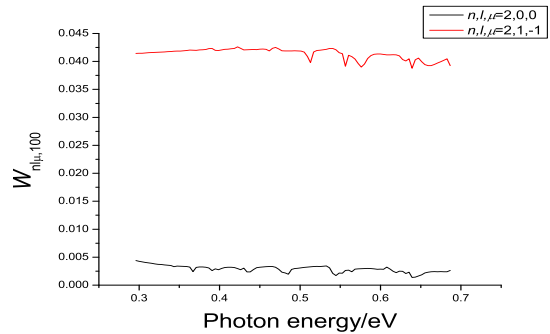,width=8cm, height=5cm}}
\caption{Transition probabilities of a hydrogen atom interacting
with a circularly polarized laser of $A=5\times10^{-6}$V$\cdot$s/m,
and their dependence on the photon energy. }\label{c1}
\end{figure}
\begin{figure}[h]
\centerline{\epsfig{file=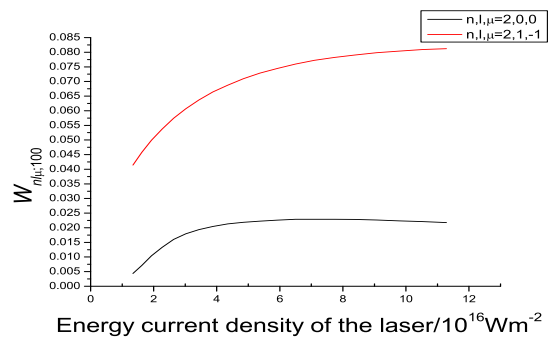,width=8cm, height=5cm}}
\caption{Transition probabilities of a hydrogen atom interacting
with a circularly polarized laser of $\hbar\omega=0.296$eV, and
their dependence on the laser intensity. }\label{c2}
\end{figure}

Fig.\ref{c1} shows that the spectrum is continuous. No discrete
sharp resonance peaks appear. If one fits the spectrum by
\begin{eqnarray}
E_{n'}-E_n=\eta \hbar\omega \ ,\label{3}
\end{eqnarray}
$\eta$ may take any real number in a wide range, not necessarily be
an integer. We call this kind of transition non-integer. It violates
the Bohr condition. When one regards the generalized Bohr condition
\begin{eqnarray}
E_{n'}-E_n=N\hbar\omega \label{02}
\end{eqnarray}
with integer $N$ as an expression of energy conservation, he has
omitted the interaction energy between the atom and the
electromagnetic field. It is permissible only when the interaction
is weak and may be handled by perturbation. The non-integer
transition is a typical non-perturbative effect, and the violation
of Bohr condition is therefore not surprising. While fig.\ref{c2}
shows that the transition probability is not proportional to an
integer power of the intensity. It means, the interaction between
the laser and the atom cannot be reduced to interactions of
individual photons with the atom separately. The interaction is
between the atom and the laser as a whole. We call this character
the non-linear effect. This scenery is quite different from the
regularity one saw in the weak light (including weak laser)
spectroscopy, therefore has to be checked by new experiments. One
may observe the radiation of the atom after it has been irradiated
by an intense laser. In this way, the changes of distributions of
atoms among various energy levels, and therefore their transition
probabilities, are measured. Although there is not any separate
resonance, fig.\ref{c1} still shows complex structure in the
spectrum. It is interesting to find out the information exposed by
this kind of structure.

\section{ Laser photo-ionizations  \label{s5}}

The photo-ionization or the photoelectric effect is the transition
of the electron from the ground state to the ionized state, when it
is irradiated by light. The photo-ionization by an intense
circularly polarized laser may be handled by the method proposed in
\cite{z}.

It is shown in \cite{z}, that the energy of the ionized electron
(photo-electron) is
\begin{eqnarray}
E_{f0}=E_i-\mu\hbar\omega\ ,\label{31}
\end{eqnarray}
and the transition probability per unit time is
\begin{eqnarray}
P=\frac{2\pi}{\hbar}\left|H_{f0,i}^{\prime\prime}\right|^2\rho\
,\label{32}
\end{eqnarray}
with
\begin{eqnarray}
H_{f0,i}^{\prime\prime}\equiv\int\psi_{f0}^*({\textit{\textbf
r}})\hat{H}^{\prime\prime}\phi_i({\textit{\textbf r}}){\rm
d}{\textit{\textbf r}}\ .\label{33}
\end{eqnarray}
$\phi_i({\textit{\textbf r}})$ in (\ref{33}) is an eigenfunction
of $\hat{H}_{\rm ps}$, satisfying (\ref{21}). $E_i$ in (\ref{31})
is the corresponding eigenvalue. $\psi_{f0}({\textit{\textbf r}})$
is the eigenfunction of $\hat{H}_0^\prime$, with eigenvalue
$E_{f0}+\mu\hbar\omega$, therefore is a projection of the Coulomb
wave function onto the subspace with definite magnetic quantum
number $\mu$, and describes the ionized electron.

In the weak light limit, $A\rightarrow 0$, $\phi_i$ approaches an
eigenfunction of $\hat{H}_0^\prime$, which is also the ground
state eigenfunction of $\hat{H}_0$ with zero magnetic quantum
number; and $E_i$ approaches the corresponding eigenvalue. They
are independent of $A$. For the hydrogen atom, they are
$\psi_{100}$ and $-b$ respectively, $b$ is the binding energy of
the electron in the ground state hydrogen atom. In the case of
$ka_0<< 1$, we have (\ref{280}), therefore the selection rule
(\ref{281}) works. These limits make the energy (\ref{31}) of the
photoelectron be
\begin{eqnarray}
E_{f0}=\hbar\omega-b\ , \label{34}
\end{eqnarray}
and the transition probability $P$ proportional to the light
intensity. This example shows, in the weak light limit, the
photo-ionization has the following distinct characters:

\vskip 0.1in \noindent C1.There is a critical frequency for a
given system.The light with frequency lower than this critical
value cannot eject any electron from the system.

\noindent C2.The light with frequency higher than this critical
value can ionize the system, the energy of the ejected electron
increases linearly with the increasing of the frequency but is
independent of the intensity of the light.

\noindent C3.The intensity of the photo-electric current is
proportional to the intensity of the light.

\vskip 0.1in \noindent This is exactly the experimental knowledge
on photo-ionization, people had before the discovery of the laser.
Based on this knowledge and guided by his idea of light quanta,
one hundred years ago, Einstein \cite{e} found his famous formula
(\ref{34}) and the idea that the light-atom interaction may be
reduced to the interactions between photons and atoms. In this
way, he explained the above experimental characters of
photo-ionization. This was a crucial step towards the discovery of
quantum mechanics. Now we see, all of these experimental
characters, as well as the Einstein formula (\ref{34}), together
with his idea that photons interact with atoms independently, are
the perturbation results of quantum mechanics in the weak light
limit. What will be the scenery when the light becomes an intense
laser?

\begin{figure}[h]
\centerline{\epsfig{file=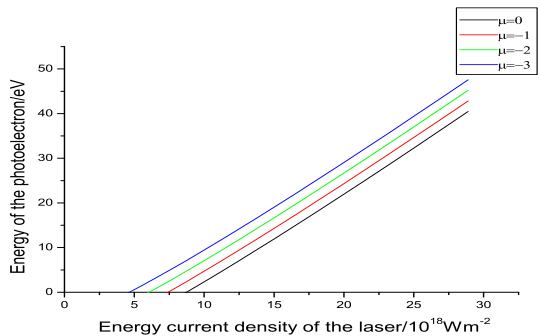,width=8cm, height=5cm}} \caption{The
energy of a photoelectron ejected from the ground state hydrogen
atom by a circularly polarized laser of $\hbar\omega=2.37$eV, and
its dependence on the laser intensity. }\label{c3}
\end{figure}

If one puts $E_1=-b$ , and $E_2=E_{f0}$, the formula (\ref{34})
becomes the Bohr condition $E_2-E_1=\hbar\omega $. Therefore, the
Einstein formula is a predecessor of the Bohr condition. Soon after
the discovery of laser, people observed MPI\cite{v} and ATI\cite{a}.
Einstein formula was generalized to be
\begin{eqnarray}
E_{f0}=N\hbar\omega-b\ ,\label{35}
\end{eqnarray}
with $N$ being an positive integer. It is a special case of the
generalized Bohr condition $ E_2-E_1=N\hbar\omega \ ,$ and may be
deduced by the higher order perturbation. (\ref{31}) would be an
exact expression of the photoelectron energy, if $E_i$ in it is
solved from (\ref{21}) exactly. Defining
\begin{eqnarray}
\eta\equiv\frac{E_i+b}{\hbar\omega}-\mu\ ,
\end{eqnarray}
one may write (\ref{31}) in the form
\begin{eqnarray}
E_{f0}=\eta\hbar\omega-b\ .\label{36}
\end{eqnarray}
It is $ E_2-E_1=\eta\hbar\omega $. Here we see, $\eta$ is an integer
only when $E_i+b-\mu\hbar\omega$ is an integer multiple of the
photon energy $\hbar\omega$. This will not necessarily be the case
for a non-perturbation interaction between the atom and an intense
laser. It means the photo-ionization may be a non-integer quantum
transition. This is a true non-perturbation effect, which cannot be
handled by perturbation of any order. Using the numerical solution
of (\ref{21}) obtained in the last section, we find the light
intensity dependence of the photoelectron energy. The numerical
result is shown in fig.\ref{c3}.

\begin{figure}[h]
\centerline{\epsfig{file=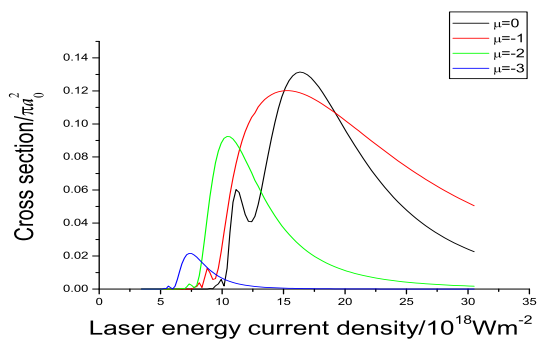,width=8cm, height=5cm}}
\caption{Cross section of the photo-ionization of a ground state
hydrogen atom irradiated by a circularly polarized laser of
$\hbar\omega=2.37$eV, and its dependence on laser intensity.
}\label{c4}
\end{figure}

The transition probability (\ref{32}) may be expressed in the form
of cross section. It is the formula (14) or (15) in \cite{z}.
Applying it to the photo-ionization of the hydrogen atom
irradiated by a circularly polarized laser, under the condition
$ka_0<<1$, we obtain the cross section
\begin{eqnarray}
\sigma=\frac{16\alpha
v}{ka_0c}\sum_{l=|\mu|}\left|\beta_l(i)\right|^2\label{37}
\end{eqnarray}
in unit of $\pi a_0^2$. Here, $v=\sqrt{2E_{f0}/m}$ is the velocity
of the photoelectron, and $\beta_l(i)$ is an elementary but some
what lengthy and tedious expression, containing integrals of the
type
\begin{eqnarray}
&&\int_0^\infty {\rm e}^{-st}t^{u-1}{\rm F}(a_1,c_1,t){\rm
F}(a_2,c_2,qt){\rm d}t\nonumber\\&&=\Gamma(u)s^{-u}{\rm
F}_2(u,a_1,a_2,c_1,c_2,s^{-1},\frac{q}{s}) \ .
\end{eqnarray}
The integral has been analytically worked out. There are two
confluent hypergeometric functions F on the left. One is from the
radial wave function of the electron in the hydrogen atom, and
another is from the Coulomb wave function of the outgoing electron.
${\rm F}_2$ on the right is the Appell's hypergeometric function of
the second class  in two variables\cite{er}. In our problem here, it
degenerates into a polynomial in two variables. Therefore the
calculation of $\beta_l(i)$ is finite, if the expansion (\ref{23})
is truncated. The calculated cross section and its dependence on the
laser intensity is shown in fig.\ref{c4}.

We see from fig.\ref{c3}, the energy of the photo-electron increases
with the increasing of the light intensity. The critical frequency
is not absolute. Even though  the frequency of the incident light is
lower than the critical frequency, the electron may still be
ejected, if the light is intense enough. The characters C2 and C1,
together with the formula (\ref{34}), are not true for intense laser
photo-ionization. Furthermore, the formula (\ref{35}) is not true
either, if the incident laser is very intense. In this case, it has
to be substituted by (\ref{36}), with possibly non-integer $\eta$.
The transition in photo-ionization becomes non-integer. However, one
may still see an apparent quantum character in fig.\ref{c3}. That
is, the energy difference between photo-electrons with different
magnetic quantum number $\mu$ is an integer multiple of the quantum
energy $\hbar\omega$, as people observed in ATI. From fig.\ref{c4}
we see, the cross sections depend on the light intensity
nonlinearly. It means that the character C3 is not true for laser
photo-ionization. The interaction between light and atoms cannot be
reduced to the independent interactions between photons and atoms.
Atoms interact with the laser as a whole. This is the non-linear
effect.

\section{Conclusion \label{s7}}
We see, a laser may not only induce MPI, but also cause non-integer
transitions, if it is strong enough. The later can only be handled
by non-perturbation method, and therefore is a true non-perturbative
effect. The non-linear effect in atom-laser interaction is also
noticeable. Some quantitative details are exposed in above sections.
The correctness of these predictions have to be finally checked by
experiments. It now calls for appropriate experiments.

Quantum electrodynamics is the best theory nowadays. It has been
checked in details around the vacuum state. However, its correctness
in presence of a intense electromagnetic wave still has to be
checked. We need reliable method to explore its solutions and
compare the results with experiments. The above method simplifies
the work considerably when the laser is circularly polarized. It may
be used for more general cases, to consider, for examples, the spin,
the relativity, as well as other atoms and molecules. \vskip 0.5cm

The work is supported by the National Nature Science Foundation of
China with grant number 10305001.

\end{document}